\documentclass[a4paper]{article}
\usepackage{interspeech2013,amssymb,amsmath,epsfig,bbm}
\usepackage{graphics,subfigure}
\ninept
\def\reg{{\rm\ooalign{\hfil
     \raise.07ex\hbox{\scriptsize R}\hfil\crcr\mathhexbox20D}}}

\title{Non-negative Matrix Factorization with Linear Constraints for Single-Channel Speech Enhancement}

\makeatletter
\def\name#1{\gdef\@name{#1\\}}
\makeatother
\name{{\em Nikolay Lyubimov$^1$, Mikhail Kotov$^2$}}

\address{$^1$Moscow State University,
  Moscow, Russia \\
  $^2$STEL - Computer Systems Ltd., Moscow, Russia\\
{\small \tt lubimov.nicolas@gmail.com, kotov@stel.ru}}

\begin{document}
\maketitle
\begin{abstract}
This paper investigates a non-negative matrix factorization (NMF)-based approach to the semi-supervised single-channel speech enhancement problem where only non-stationary additive noise signals are given. The proposed method relies on sinusoidal model of speech production which is integrated inside NMF framework using linear constraints on dictionary atoms. This method is further developed to regularize harmonic amplitudes. Simple multiplicative algorithms are presented. The experimental evaluation was made on TIMIT corpus mixed with various types of noise. It has been shown that the proposed method outperforms some of the state-of-the-art noise suppression techniques in terms of signal-to-noise ratio.
\end{abstract}

\section{Introduction}
Speech enhancement in presence of background noise is an important problem that exists for a long time and still is widely studied nowadays. The efficient single-channel noise suppression (or noise reduction) techniques are essential for increasing quality and intelligibility of speech, as well as improving noise robustness for automatic speech recognition (ASR) systems.

The noise reduction mechanism is designed to eliminate additive noise of any origin from noisy speech. Most often the estimation of speech and noise signal parameters is performed in spectral domain. The successful example is the Minimum Mean Square Error (MMSE) estimator, which utilizes statistical modeling of speech and noise frequency components \cite{Ephraim84}. One of the most simple and popular approach is spectral subtraction \cite{Boll79}, where noise spectral profiles are estimated during non-speech segments, and then subtracted from speech segments in magnitude domain. However this method is restrictive to the quasi-stationarity of the observed noise.

Some methods are based on semi-supervised noise reduction, where the target noise signals are presented to train the system. Recently Non-negative Matrix Factorization (NMF) had become a widely used tool for making useful audio representations, especially in context of blind source separation \cite{Schmidt06} and semi-supervised noise reduction \cite{Schmidt07}. The latest is considered to be a special case of separation problem with additive background noise as secondary source. NMF factorizes the given magnitude spectrogram $Y$ into two non-negative matrices $Y=DX$ where columns of dictionary $D$ are magnitude spectral profiles, and $X$ contains gain coefficients. The key steps of semi-supervised NMF-based speech enhancement are 1) estimation of noise dictionary $D_n$ from the given training sample, then 2) factorizing input spectrogram using speech and noise dictionaries, i.e. $Y = D_sX_s + D_nX_n$ 
and finally 3) reconstruct using clean speech magnitude estimation $Y_s = D_sX_s$ or by using it to filter the initial complex spectrogram. Though these simple steps provide good results while reducing non-stationary noise sources \cite{Cauchi12}, they don't consider any additional knowledge about specific signal structure, lacking the interpretability of estimated components and degrading the final performance. Several methods have been proposed to overcome these limitations by introducing regularization probabilistic priors \cite{Wilson08} and constrained NMF procedure for noise estimation \cite{Mohammadiha11}. The promised results have been achieved by using non-negative hidden markov model framework\cite{Mysore11} that can effectively handle temporal dynamics. In \cite{Weninger12} the regularized NMF problem for speech and music separation is stated. This method uses normalized sparsity constraints on musical dictionary and gain factors for better discriminating between signals.

However much fewer papers are dedicated to building appropriate constraints on target speech dictionary $D_s$ itself, rather than other NMF components. Most natural constraints should enforce harmonicity of columns. In this paper we adopt the approach proposed by \cite{Bertin09} in polyphonic music transcription task and further develop it to enhance noisy speech signals.

The rest of the paper is organized as follows. The first section describes the underlying speech and noise models, which lead to NMF representation with linear constraints of dictionary columns. The multiplicative updates algorithm is applied to estimate unknown parameters. Then the regularized extension of this model is proposed. In the next section experimental evaluations on noisy TIMIT corpus are given. In conclusion we discuss about advantages and drawbacks of contributed method.

\section{Method}
This section considers speech $D_s$ and noise $D_n$ matrix dictionaries in which columns, or \textit{atoms} follow specific linear models that are discussed below.
\begin{figure*}[t]
\centering
\subfigure[]{{\epsfig{figure=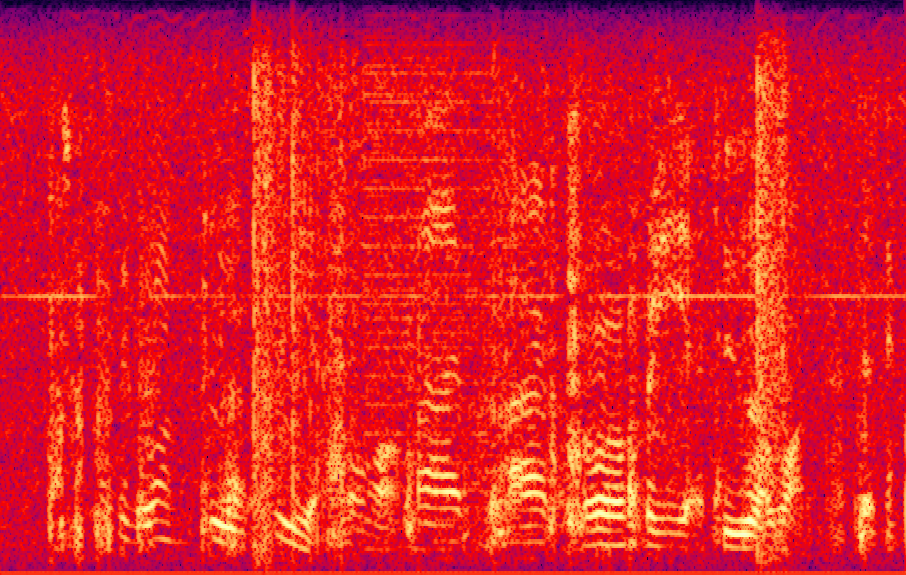, width=40mm}}}\quad
\subfigure[]{{\epsfig{figure=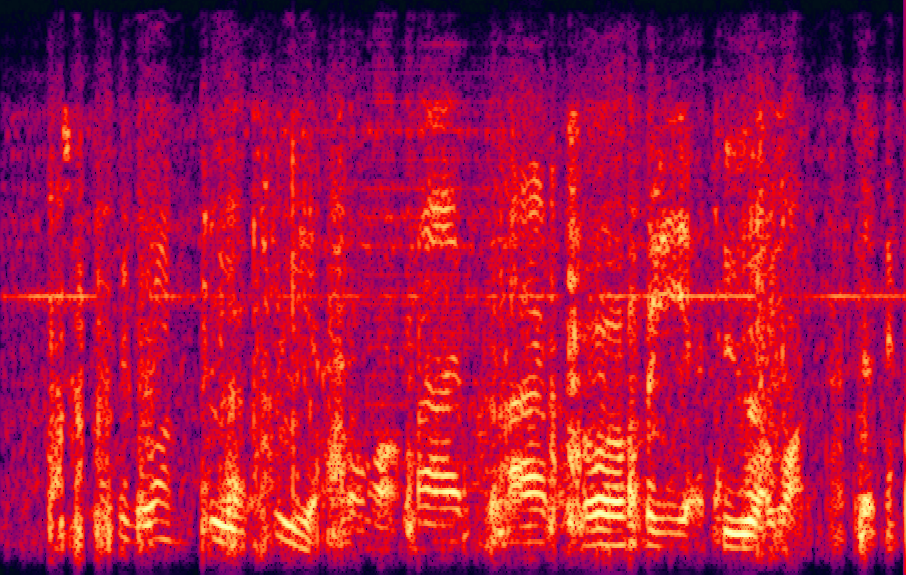, width=40mm}}}\quad
\subfigure[]{{\epsfig{figure=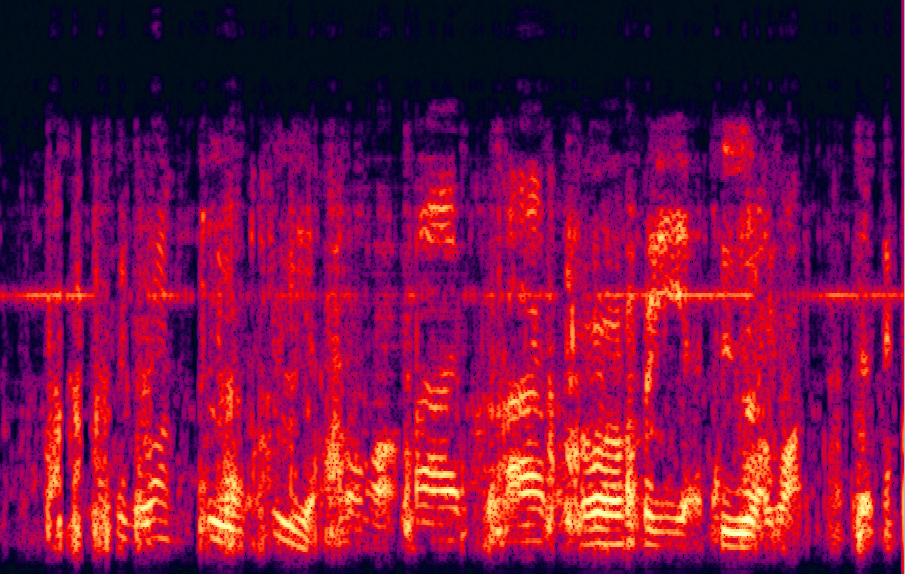, width=40mm}}}\quad
\subfigure[]{{\epsfig{figure=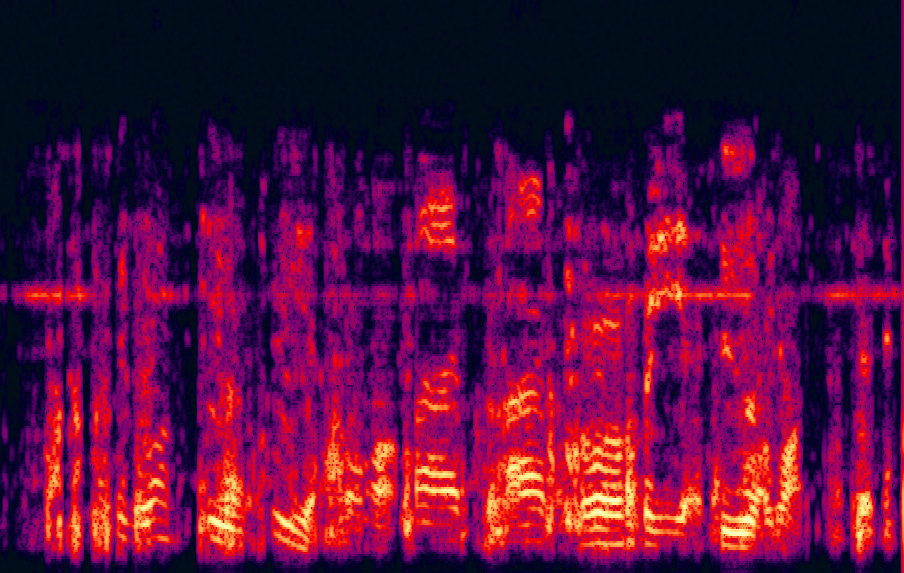, width=40mm}}}\\
\caption{{\it Spectrograms of (a) noisy signal, and denoised variant using (b) linNMF, (c) denseNMF and (d) denseNMF with higher sparsity constraints}}
\label{fig:spectrograms}
\end{figure*}
\subsection{Speech model}
The basic of speech production assumes that the time-domain excitation source $e(t)$ and vocal tract filter $a(t)$ are combined into convolutive model \cite{Gold00}:
\begin{equation}\label{eq:source-filter-model}
s(t) = a(t)*e(t)
\end{equation}
The excitation signal $e(t)$ itself could be presented by summing up complex sinusoids and noise on frequencies that are multiples of fundamental frequency \cite{McAulay86}:
\begin{equation}\label{eq:sinusoids-and-noise}
e(t) = \sum_{k=1}^p {c_k \exp{(i k \overline{\omega}(t))}}
\end{equation}
The quasi-stationarity of the transfer function $a(t)$ and fundamental frequency function $\overline{\omega}(t)$ takes place during the short frame $t \in \mathcal{T}_\tau$, so that $a(t) = a_\tau(t)$ and $\overline{\omega}(t) = \overline{\omega}_\tau$. Putting together two representations \eqref{eq:source-filter-model} and \eqref{eq:sinusoids-and-noise} provides:
\begin{equation}\label{eq:combination}
s_\tau(t) = \sum_{k=1}^p {c_k \hat{a}_\tau(k\overline{\omega}_\tau)\exp{( ik\overline{\omega}_\tau t)}}
\end{equation}
where the hat symbol indicates Fourier transform of the corresponding function. The approximation of signal \eqref{eq:combination} in magnitude Short-Time Fourier Transform (STFT) domain using window function $w(t)$ is given as:
\begin{equation}\label{eq:in_magnitude}
|\hat{s}_\tau(\omega)| \approx \sum_{k=1}^p {c_k |\hat{a}_\tau(k\overline{\omega}_\tau)||\hat{w}(\omega-k\overline{\omega}_\tau)|}
\end{equation}
since $\hat{w}(\omega)$ localizes most energy at low frequencies. Then we suppose that frequency response $|\hat{a}(\omega)|$ at any frame $\tau$ is modeled by the (possibly sparse) combination of fixed spectral shapes with non-negative weights $x_j^s(\tau)\geq0$, i.e. $|\hat{a}_\tau(\omega)| = \sum_{j=1}^m{x_j^s(\tau)|\hat{a}_j(\omega)|}$. In this case the final representation formula is:
\begin{equation}
|\hat{s}_\tau(\omega)|\approx\sum_{j=1}^m {x_j^s(\tau) \sum_{k=1}^p {c_k|\hat{a}_j(k\overline{\omega}_\tau)||\hat{w}(\omega-k\overline{\omega}_\tau)|}}
\end{equation}
The latest representation could be efficiently wrapped in matrix notation by taking discrete values $\{\omega_i\}_{i=1}^K$ and $\{\tau_l\}_{l=1}^T$, since only one fundamental frequency $\overline{\omega}_\tau$ is expected for the given time frame. It is possible to choose it from the discrete set $\overline{\Omega}_L$ of $L$ hypothesized fundamental frequencies bounded by $\overline{\omega}_{\min}$ and $\overline{\omega}_{\max}$. This leads to $m_s=Lm$ possible combinations. Therefore the following equation for the input speech spectrogram $Y_s \in \mathbb{R}_{+}^{K \times T}$ holds:
\begin{equation}\label{eq:linNMF-speech}
\begin{cases}
Y_s = D_sX_s\\
d_j^s= W_jCa_j, \quad j=1,2,\ldots m_s
\end{cases}
\end{equation}
Each column $d_j^s$ of matrix $D_s \in \mathbb{R}_{+}^{K \times m_s}$ represents one \textit{harmonic atom}, in which isolated harmonics are placed in columns of matrices $W_j \in \mathbb{R}_{+}^{K \times p}$, weighted by constant amplitude matrix $C = diag(c_1, \ldots c_p)$. The representation coefficients $a_j \in \mathbb{R}_{+}^p$ and gain matrix $X_s$ are needed to be defined.
\subsection{Noise model}
Being not so physically motivated as speech model the noise model of signal is constructed in the similar way by assuming additivity of corresponding spectral shapes (however this could also be  theoretically approved by exploring band-limited noise signals, that is not the case of the current work). Each time-varying noise magnitude $|\hat{n}_\tau(\omega)|$ is decomposed into sum of individual static components:
\begin{equation}
|\hat{n}_\tau(\omega)| = \sum_{k=1}^{r}{b_k(\tau)|\hat{n}_k(\omega)|}
\end{equation}
Note that $|\hat{n}_k(\omega)|$ is the predefined set of noise magnitudes extracted from the known noise signals, whereas non-negative filter gains $b_k(\tau)\geq0$ should be defined from the observed data. As stated before the non-negative combination $b_k^n(\tau)=\sum_{j=1}^{m_n}{x_j^n(\tau)b_{kj}}$ leads to the following representation:
\begin{equation}
|\hat{n}_\tau(\omega)| = \sum_{j=1}^{m_n}{x_j^n(\tau)\sum_{k=1}^{r}{|\hat{n}_k(\omega)|b_{kj}}}
\end{equation}
and in matrix notation for the same discrete $\omega_i$ and $\tau_l$ on input noise spectrogram $Y_n \in \mathbb{R}_{+}^{K \times T}$:
\begin{equation}\label{eq:linNMF-noise}
\begin{cases}
Y_n = D_nX_n\\
d_j^n = Nb_j, \quad j = 1,2,\ldots m_n
\end{cases}
\end{equation}
with $D_n \in \mathbb{R}_{+}^{K \times m_n}$ represents noise dictionary with atoms $d_j^n$,  $N \in \mathbb{R}_{+}^{K \times r}$ contains noise spectral shapes combined with unknown coefficients $b_j \in \mathbb{R}_{+}^{r}$ to produce noise model.
\subsection{NMF with linear constraints}
Here we introduce the general formulation of NMF problem with linear constraints that follows from speech \eqref{eq:linNMF-speech} and noise \eqref{eq:linNMF-noise} representations. As soon as spectrograms in both cases factorize by the product of two non-negative matrices, the $\beta$-divergence between observed spectrogram $Y$ and $DX$ product could be chosen for approximation \cite{Fevotte10}. Here we only consider a special case called Kullback-Leibler divergence $D_{KL}(Y\|DX)=\sum_{i,j} {Y_{ij}\log{\frac{Y_{ij}}{(DX)_{ij}}} - Y_{ij} + (DX)_{ij}}$, but other divergences could be used as well.

The following general optimization problem (\textit{linNMF}) gives solution to \eqref{eq:linNMF-speech} and \eqref{eq:linNMF-noise}:
\begin{equation}\label{eq:linNMF-task}
\begin{cases}
D_{KL}(Y\|DX) + \lambda \|X\|_1 \rightarrow \min\\
d_j = \Psi_j a_j, \qquad j=1,2,\ldots m
\end{cases}
\end{equation}
with minimization over $\{a_j\}_{j=1}^m$ and $X$. The factors $d_j$ are spanned by the columns of matrices $\Psi_j$ that could be diverse for $j=1,2\ldots m$. Using multiplicative updates heuristic \cite{Fevotte10}, it could be shown that the following algorithm solves the optimization problem \eqref{eq:linNMF-task}:
\begin{equation}\label{eq:linNMF-updates}
\begin{split}
a_j &\longleftarrow a_j \cdot \dfrac{\Psi_j^T\frac{Y}{DX}\overline{x}_j^T}{\Psi_j^T\mathbf{1}\overline{x}_j^T}\\
X &\longleftarrow X \cdot \dfrac{D^T\frac{Y}{DX}}{(D^T\mathbf{1} + \lambda)}
\end{split}
\end{equation}
where $\overline{x}_j$ denotes $j$-th row of matrix $X$, $\mathbf{1} \in \mathbb{R}^{K \times T}$ is the all-ones matrix and multiplications $a\cdot b$ and divisions $\frac{x}{y}$ are element-wise. By iterating these rules the factors $d_j$ are modeled in linear subspace with dimensionality implied by rank of $\Psi_j$.
\subsection{NMF with linear dense constraints}
It has been experimentally found that many solutions achieved by \eqref{eq:linNMF-updates} tend to ''reduce'' the rank of corresponding subspace. In other words, each $d_j$ tends to have sparse representation in basis $\Psi_j$. It is not the desired solution in the current task, in case if $\Psi_j$ contains windowed-sinusoid magnitude values on the particular frequency. As we want to extract full harmonic atoms from signal, it is expected that every harmonic has non-zero amplitude (especially in low-frequency band). The \textit{denseNMF} optimization task is proposed to overcome this problem that favors non-zero coefficients in vectors $\{a_j\}_{j=1}^m$:
\begin{equation}\label{eq:denseNMF-task}
\begin{cases}
D_{KL}(Y\|DX) + \lambda \|X\|_1 + \alpha \sum_j{\|a_j\|_2^2} \rightarrow \min \\
d_j = \Psi_j a_j, \, \|a_j\|_1 = 1\qquad j=1,2,\ldots m
\end{cases}
\end{equation}
The following rules are also derived from multiplicative updates for $l_1$-normalized coefficients $\tilde{a}_j = a_j/\|a_j\|_1$:
\begin{equation}\label{eq:denseNMF-updates}
\begin{split}
a_j &\longleftarrow \tilde{a}_j \cdot \dfrac{\mathbbm{1}_j\tilde{a}_j^T\Psi_j^T\mathbf{1}\overline{x}_j^T + \Psi_j^T\frac{Y}{DX}\overline{x}_j^T + \alpha\mathbbm{1}_j\tilde{a}_j^T\tilde{a}_j}{\Psi_j^T\mathbf{1}\overline{x}_j^T + \mathbbm{1}_j\tilde{a}_j^T \Psi_j^T\frac{Y}{DX}\overline{x}_j^T + \alpha\tilde{a}_j}\\
X &\longleftarrow X \cdot \dfrac{D^T\frac{Y}{DX}}{(D^T\mathbf{1} + \lambda)}
\end{split}
\end{equation}
where $\mathbbm{1}_j$ indicates the vector of all-ones of the same size as $a_j$. It should be noted that convergence properties of presented algorithms \eqref{eq:linNMF-updates}\eqref{eq:denseNMF-updates} are not studied. However during the experiments the monotonic behavior of the target function \eqref{eq:linNMF-task} has been permanently observed for arbitrary $\alpha>0$.

\subsection{Speech enhancement}\label{sseq:speech_enhancement}
The actual speech enhancement follows the similar procedure as described in \cite{Schmidt07} and \cite{Cauchi12}. The initial noise shapes $N$ are preliminary extracted from the noise-only magnitude spectrograms. These shapes are used directly to form noise atoms as in \eqref{eq:linNMF-noise}. 

Harmonic atoms are constructed using Fourier transform of Hann window shifted on $k\overline{\omega}$ and scaled by $c_k = Sinc^2(\frac{k\overline{\omega}}{2})$ that approximates triangle pulse excitation. This process requires the following parameters to be set:
\begin{itemize}
\item the set of hypothesized fundamental frequencies $\overline{\Omega}_L$ that consists of frequency limits $\overline{\omega}_{\min}$ and $\overline{\omega}_{\max}$, and $L$ points equally spaced between these limits. As we are going to show in experimental evaluations, $L$ parameter has drastic effect on noise suppression performance.
\item the number of harmonic atoms $m$ to be estimated per each hypothesized fundamental frequency.
\item the number of harmonics $p$. For each atom $p=\min\{p^*, \lfloor\frac{sr}{2\overline{\omega}}\rfloor\}$ is chosen, where $sr$ is the sampling rate.
\end{itemize}

The constrained speech and noise dictionaries $D_s$ and $D_n$ are then combined into matrix $D = [D_s D_n]$, and gain matrix $X = [X_s^T X_n^T]^T$ is randomly initialized. Using iterative updates \eqref{eq:linNMF-updates} or \eqref{eq:denseNMF-updates} all representation coefficients and gain matrices are obtained using different sparsity parameters $\lambda_s$ and $\lambda_n$ for speech and noise components\cite{Schmidt07}\cite{Cauchi12}. Finally the denoised signal is made via inverse STFT of Wiener-filtered input (see \cite{Cauchi12} for details).

\section{Experimental results}
\begin{figure*}[ht!]
\subfigure[Input SNR = -5 dB]{\epsfig{figure=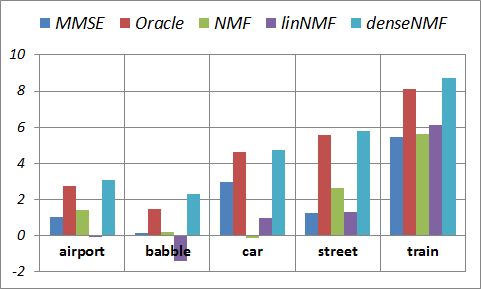, width=80mm}}
\subfigure[Input SNR = 0 dB]{\epsfig{figure=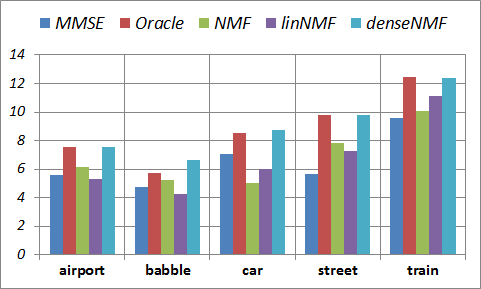, width=80mm}}\\
\subfigure[Input SNR = 5 dB]{\epsfig{figure=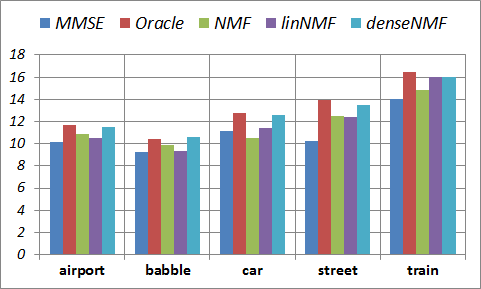, width=80mm}}
\subfigure[Input SNR = 15 dB]{\epsfig{figure=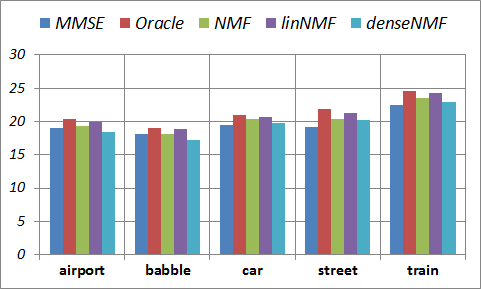, width=80mm}}
\caption{{\it Performance evaluations (in terms of output SNR w.r.t different noisy environments) on two proposed algorithms (\textit{linNMF}, \textit{denseNMF}) and other baseline speech enhancement methods (MMSE\cite{Ephraim84}, NMF\cite{Schmidt07}\cite{Cauchi12}) and Oracle estimator with known clean signals.}}
\label{fig:SNR-on-TIMIT}
\end{figure*}
Figure \ref{fig:spectrograms} illustrates the denoised spectrograms of short excerpt of speech contaminated by non-stationary noise. To demonstrate the performance we have tested the proposed NMF-based algorithms on TIMIT corpus \cite{TIMIT}. 1245 sentences from 249 speakers were mixed with various types of non-stationary noises, taken from airport, car, train, babble and street backgrounds. We have used signal-to-noise ratio (SNR) as objective quality measure \cite{Hu08}. Four types of input SNR are explored: -5dB, 0dB, 5dB, 15dB.

The algorithms parameter settings are the following. The sampling rate of all signals is $sr$ = 8KHz. The STFT was applied to noise and noisy signals with 32 ms Hanning window and 75\% overlap. Then noise matrix $N$ with $r$ = 16 columns is trained using non-constrained NMF on 10 seconds noise spectrogram of known type of noise. All atoms $\Psi_j$ are constructed using the following parameters (see section \ref{sseq:speech_enhancement}): $\overline{\omega}_{\min}$ = $\frac{2\pi}{sr}$80Hz, $\overline{\omega}_{\max}$ = $\frac{2\pi}{sr}$400Hz, $L$ = 33, $m$ = 4, and $p^*$ = 30 for harmonic atoms, and $m_n$ = 16 for noise atoms. Then \textit{linNMF} and \textit{denseNMF} were applied to input noisy magnitude by doing \eqref{eq:linNMF-updates} and \eqref{eq:denseNMF-updates} by fixing the number of iterations to 25. $\lambda_s$ = 0.2 sparsity coefficient was chosen for speech component, and $\lambda_n$ = 0 for noise component according to \cite{Cauchi12}. For \textit{denseNMF} algorithm the regularization parameter $\alpha$ was set to $10$, that was found experimentally.

Besides proposed algorithms we have also included two over NMF-based algorithms that serve as baseline. The first one uses known speech dictionary $D_s$ estimated directly from the same clean signal. Then NMF computes only gain matrices $X_s$ and $X_n$, followed by the same reconstruction. This is called \textit{Oracle} speech enhancement in this evaluations. The second one, called here \textit{NMF}, is also NMF-based noise reduction with unknown speech dictionary $D_s$ which is similar to those described in \cite{Schmidt07} and \cite{Cauchi12}. Also \textit{MMSE} technique \cite{Ephraim84} has been included as reference, though this algorithm is completely blind.

The figure \ref{fig:SNR-on-TIMIT} demonstrates the SNR performance of proposed algorithms. It is shown that for low SNR \textit{denseNMF} gives optimistic results, outperforming sometimes even ''\textit{oracle}'' evaluations. However it is not so surprising because oracle algorithm actually doesn't contain any intrinsic speech or noise model,poorly discriminating between both signals. For higher SNR \textit{denseNMF} starts to perform worse than proposed \textit{linNMF} and even than baseline algorithms. It probably happens due to underestimation of non-harmonic speech components.
\begin{figure}[t]
\centerline{\epsfig{figure=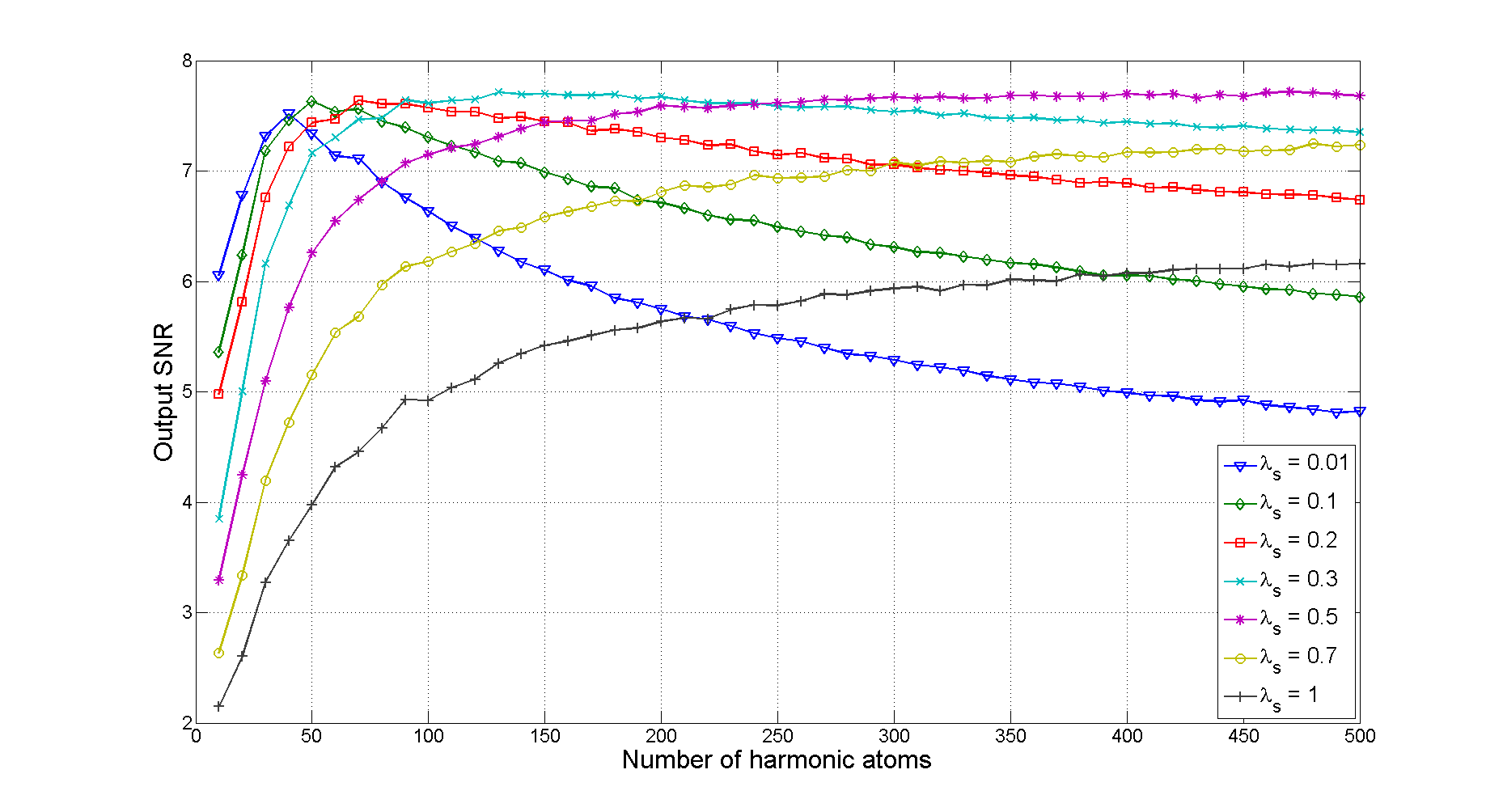,width=80mm}}
\caption{{\it Comparison of denseNMF output SNR performance for different sparsity constraints w.r.t the number of harmonic atoms used in speech dictionary $D_s$}}
\label{fig:num-vs-sparsity}
\end{figure}

Finally we have studied how the number of harmonic atoms ($m_s = Lm$) and different values of sparsity parameter $\lambda_s$ affect \textit{denseNMF} noise suppression performance. Ranging $L$ from 2 to 100 with $m=5$, it has resulted to 10 to 500 atoms in total. We have expected that higher number of atoms would give better results for high sparsity and worse for low sparsity, since the high number of sinusoids tend to overestimate the presented noise. This expectations are confirmed by the results depicted on figure \ref{fig:num-vs-sparsity}. Indeed, for $\lambda_s=0.5$ the maximum appears in more dense mixture of harmonic atoms, whereas lesser values of $\lambda_s$ have peaks at $L$ around $30$, and then rapidly decrease when more atoms are involved. Note that though setting $\lambda_s=1$ seems to degrade the overall performance, actually we observed that using this sparsity and high number of atoms gives more audibly pleasant result with much less noise but bit more distorted harmonics, as also could be observed from figure \ref{fig:spectrograms}d. Hence other objective quality metrics such as in \cite{Hu08} should be investigated in the future works.
\section{Conclusions}
The main contributions of this paper are following. First we have applied prior deterministic modeling of speech and noise signals inside NMF-based speech enhancement framework. It has led to linearly constrained columns or atoms of corresponding dictionaries $D_s$ and $D_n$. Then we have proposed the new optimization problem statement and multiplicative updates algorithm that regularizes the representation coefficients so that they contain as fewer zeros as possible, i.e., acquiring dense solution. We have tested the new method on TIMIT corpus mixed with noises on different SNR, achieving the best result for low SNR among some state-of-the-art noise suppression algorithms \cite{Ephraim84}\cite{Schmidt07}, and slightly outperforming ''\textit{oracle}'' NMF-estimator with known clean signals.

However, despite the proposed NMF regularizer, most of the extracted speech atoms are still not purely harmonic. It is a complicated problem in order to develop more efficient speech enhancement method, as well as using these atoms for further analysis. Therefore we will try to put some other NMF constraints that enforce spectral envelope smoothness in time and frequency to achieve more reliable results in future works.

\newpage
\eightpt
\bibliographystyle{IEEEtran}

\end{document}